\documentclass[twocolumn]{pasj02}
\usepackage[utf8]{inputenc}
\usepackage{graphicx}
\usepackage{multirow}
\usepackage{xcolor}
\usepackage{caption}
\usepackage{comment}
\usepackage[switch,mathlines]{lineno}
\usepackage{amstext}
\jyear{2024}
\Received{}
\Accepted{}

\begin{document}

\title{Fast variability and circular polarization of the 6.7 GHz methanol maser in G33.641$-$0.228}

\author{Kenta \textsc{fujisawa}\altaffilmark{1}}
\author{Yui \textsc{sugiura}\altaffilmark{2}}
\author{Yuta \textsc{kojima}\altaffilmark{3}}
\author{Koichiro \textsc{sugiyama}\altaffilmark{4,1}}
\author{Kotaro \textsc{niinuma}\altaffilmark{3}}
\author{Kazuhito \textsc{motogi}\altaffilmark{3}}
\author{Yoshihiro \textsc{tanabe}\altaffilmark{5,1}}
\author{Yoshinori \textsc{yonekura}\altaffilmark{5}}

\altaffiltext{1}
{Research Institute for Time Studies, Yamaguchi University, 1677-1 Yoshida, Yamaguchi-city, Yamaguchi 753-8511, Japan}
\altaffiltext{2}
{Faculy of Science, Yamaguchi University, 1677-1 Yoshida, Yamaguchi-city, Yamaguchi 753-8512, Japan}
\altaffiltext{3}
{Graduate school of Science and Technology for Innovation, Yamaguchi University, 1677-1 Yoshida, Yamaguchi-city, Yamaguchi 753-8512, Japan}
\altaffiltext{4}
{National Astronomical Research Institute of Thailand (Public Organization), 260 Moo 4, T. Donkaew, A. Maerim, Chiang Mai, 50180, Thailand}
\altaffiltext{5}
{Center for Astronomy, Ibaraki University, 2-1-1 Bunkyo, Mito, Ibaraki 310-8512, Japan}

\email{kenta@yamaguchi-u.ac.jp}

\maketitle

\begin{abstract}

The 6.7 GHz methanol maser in a high-mass star-forming region G33.641$-$0.228 is known to exhibit burst-like flux variability due to an unknown mechanism.
To investigate the burst mechanism, we conducted high-cadence flux and circular polarization monitoring observations, simultaneously using left- and right-hand circular polarizations. 
We found that the flux density increased and decreased on a short timescale of 0.3 d during a bursts. We also found strong circular polarization, reaching up to 20\% in the component exhibiting the bursts. 
Circular polarization of 0--20\% was continuously observed from 2009 to 2016, even in the quiescent period. 
The polarization also varied on timescales of less than one day. When bursts occurred and the flux density increased, the circular polarization decreased to zero. 
To explain the observational properties of the flux variability and circular polarization, we propose a model in which an explosive event similar to a solar radio burst occurs on the line of sight behind the maser cloud, producing circularly polarized continuum emission due to gyro-synchrotron or gyro-resonance radiation, which is then amplified by the maser.
\end{abstract}


\section{Introduction}
The 6.7 GHz methanol maser is emitted from high-mass star-forming regions and exhibits several interesting features in its flux variability. 
Periodic variability was reported by Goedhart et al.\ (2004), which was the first report of periodic variability of any kind in high-mass star-forming regions. Sugiyama et al.\ (2008) and Szymczak et al.\ (2014) found that variability of multiple spectral components of the 6.7 GHz methanol maser in Cep A showed correlation and anti-correlation among the components. 
In accretion bursts, a temporary rapid accretion of gas from the disk toward the central star occurs, resulting in an increase in infrared and maser luminosity (Fujisawa et al.\ 2015, Meyer et al.\ 2017, Burns et al.\ 2020). 
These are variabilities on timescales of about one month to one year. 
Regarding short-term flux variability, the discovery of the shortest periodic variability (P $=$ 23.9 d) in G14.23$-$00.50 by Sugiyama et al.\ (2017) and flares occurring on a timescale of several days in G107.298$+$5.639 (Fujisawa et al.\ 2014a, Szymczak et al.\ 2016, Aberfelds et al.\ 2023) has been reported.\footnote{Tanabe et al. (in preparation) have identified a periodic methanol maser source with a period of 21.9 d.} 
It is widely accepted that the flux variability of the 6.7 GHz methanol maser is mainly caused by changes in the excitation state of the methanol molecules due to the variability of the luminosity of the exciting source, but it has also been pointed out that flux variability in background radiation can be the cause of maser variability (e.g., Araya et al.\ 2010, Szymczak et al.\ 2011, Olech et al.\ 2020, Gray et al.\ 2020).

The maser burst contrasting with these variabilities has been reported in the 6.7 GHz methanol maser in G33.641$-$0.228 (Fujisawa et al. 2012, 2014b).
G33.641$-$0.228 is a high-mass star-forming region with a kinematic distance of 7.6 kpc (Reid et al.\ 2019) and an infrared luminosity based on the IRAS database of 4.4$\times$10$^4\ \LO$.
The 6.7 GHz methanol maser in this source was first reported by Szymczak et al.\ (2000), showing multiple peaks (hereafter referred to as components). One (non-bursting) component exhibiting periodic variability of about 500 days, as reported by Olech et al.\ (2019). 
VLBI observations have revealed that the spots of the 6.7 GHz methanol maser are distributed elliptically (Fujisawa et al.\ 2012; Bartkiewicz et al.\ 2009) and that water maser spots are distributed perpendicular to it (Bartkiewicz et al.\ 2011, 2012). 

The characteristics of the burst in G33.641$-$0.228 can be summarized as follows: only one of the multiple spectral components (Component II, $V_{\rm LSR}$ $=$ 59.6 km s$^{-1}$) shows the burst, while the others show no changes during the burst. 
The flux density increases significantly (e.g., seven times in Fujisawa et al.\ 2012) on a timescale of less than one day when the burst occurs, and then decreases on a timescale of about five days after the burst. 
The decrease is not monotonic and often shows fast fluctuation (Fujisawa et al.\ 2012, 2014b; Berzins et al.\ 2018). 
No periodicity has been reported in the occurrence of the burst so far. Szymczak et al.\ (2018) reported the discovery of similar short-term burst phenomena in five sources.
The mechanism underlying the burst in G33.641$-$0.228 is currently unknown. 
To explain the short-term variability (timescale of less than 1 day), Fujisawa et al.\ (2012, 2014b) proposed explosive local heating, while Berzins et al.\ (2018) suggested the possibility of thin, collimated outbursts emitted by the central star causing local variations in magnetic field strength based on observations showing flux density repeating short-term increases and decreases during the post-burst decrease phase. Khaibrakhmanov et al. (2025) proposed a physical model in which the reconnection between the circumstellar magnetic field and the disk magnetic field releases magnetic energy in short time, causing local heating and changing the excitation state of the maser, resulting in a maser burst.

The light curve during a burst of G33.641$-$0.228 resembles that of solar radio bursts; solar radio burst occurs when energy stored in the magnetic field is suddenly released (e.g., Dulk\ 1985). Gyro-synchrotron radiation is produced by the interaction of accelerated electrons with the magnetic field, and the emission is often circularly polarized. The typical timescale of solar radio bursts is 10 minutes in the microwave band. A rapid increase in flux density followed by a decrease with a somewhat longer timescale. The flux variability often involves oscillatory fluctuations. 

Similar mechanism may be involved in the burst of G33.641$-$0.228. If gyro-synchrotron radiation is involved, short-term flux variability and circularly polarized maser emission may occur. So far, monitoring observations have been performed at a cadence of only about once a day, and no detailed light curves during bursts have not been reported. Moreover, observations of circular polarization have not been reported.

We investigate the mechanism of the burst focusing on the detail light-curve and circular polarization. We conducted a single-dish monitoring observation at high cadence during a burst, and analyzed the data of right and left circular polarization (RHCP and LHCP) independently as well as reanalyzing the past observation data. In this paper, we present the observations in Chapter 2, the results in Chapter 3, the discussion in Chapter 4, and the conclusion in Chapter 5.

\section{Observations}
The observations were conducted using the Yamaguchi 32-meter radio telescope (Fujisawa et al.\ 2002). 
The dates of observation are summarized in Table \ref{obstab}.

\begin{table*}[h!]
  \begin{center}
\caption{Observation date}
  \begin{tabular}{ccccc}
  \hline
 Year&Date&MJD&Observation period\footnotemark[$*$]&Number of scan\footnotemark[$\dag$]\\
  &&& (day)&\\\hline
2009&July 2--October 14&55014--55121&108&61\\
2010&July 15--September 23&55392--55462&71&106\\
2011&September 12--November 24&55816--55889&74&41\\
2012&September 28--November 11&56198--56242&45&40\\
2014--2015&June 19--January 25 (2015)&56827--57047&221&606\\
2016&January 2--10&57389--57397&9&291\\\hline
  \end{tabular}
  \label{obstab}
    \end{center}
  \begin{tabnote}
  \footnotemark[$*$]The number of days from the beginning to the end of the observation period.\\
  \footnotemark[$\dag$]The number of obtained data, including multiple observations on one day.
  \end{tabnote}
  \label{tb1}  
\end{table*}

The first four observations were conducted for short periods in each of the years 2009, 2010, 2011, and 2012, and the resulting light curves were reported in Fujisawa et al.\ (2014b). One scan was performed per day in the observations, but in 2010, there were days when multiple scans were conducted. Fujisawa et al.\ (2012, 2014b) reported the averaged spectra of both left and right circular polarization. In this study we reanalyzed the data separately for left and right circular polarization.

The observations during 2014--2015 were conducted for a total of 221 d, from June 19, 2014 (MJD 56827), to January 25, 2015 (MJD 57047). Normally, two scans were conducted per day during the observation period. A large burst occurred on MJD 56898 (August 29, 2014). After the burst, high-cadence observations were conducted for 23 days (MJD 56899--56921) for 7 hours each day (13 scans per day) starting from the day after the burst (we call this campaign as `2014 burst'). Similar continuous observations were conducted at the last part of 2014--2015 observation for 21 days in 2015 (2015 intensive monitoring). No burst occurred during 2015 intensive monitoring. To measure the flux density accurately in the 2014--2015 observation, pointing errors were corrected for each scan. After these corrections, there were still systematic errors that could not be fully corrected, e.g., the observed flux density variability showed the same pattern every day, which indicates that the correction for the elevation dependence of the gain was insufficient. The measured flux densities have a systematic error of up to 15\%.

The last session was conducted everyday from January 2 to 10, 2016 (2016 intensive monitoring), excluding January 4 and 5. The observation time per day was 6 to 8 hours, and the number of acquired data was 36 to 48 per day. Pointing correction of each scan was not performed, and G33.641$-$0.228 was continuously observed from the beginning to the end of each day. 

The observation system and the data reduction process were basically the same as those described in Fujisawa et al.\ (2012, 2014b), except for the separate analysis of circular polarization. The frequency and velocity resolutions of the spectrum were 0.976 kHz and 0.0439 km s$^{-1}$, respectively. The integration time was 840 seconds, but it was 595 seconds for the observation in 2016. Prior to 2012, a room-temperature receiver with a system noise temperature of 200 K was used, while after 2014, a cooled receiver was used, and the typical system noise temperature was 50 K but increased up to 80 K in rainy conditions. The 1$\sigma$ noise levels are 1.9 Jy (2009--2012), 0.47 Jy (2014--2015), and 0.56 Jy (2016). 
Observations were carried out by right and left circular polarizations using a circular polarizer with a cross-polarization separation of 20 dB or higher. The contamination of two polarization data due to imperfect cross-polarization separation is small enough, and is ignored in the following discussion.

The data of left and right circular polarizations were reduced independently after the polarizer, so the accuracy of the calibration of the absolute value of the flux density limited the accuracy of polarization measurement, and it was difficult to directly measure the small circular polarization as an absolute value. We assumed that spectral component III, which has a peak near Vlsr = 60.3 km s$^{-1}$, is unpolarized (circular polarization $=$ 0), and performed all calibrations and analyses based on this assumption. With this assumption, the polarization of spectra other than component III is measured relative to it, and the accuracy of the circular polarization increases to the level of the thermal noise. In section 4.1, we discuss the validity of the assumption that component III is unpolarized.

The flux density of each component in both left- and right-hand circular polarizations was derived as follows. First, Gaussian fitting was applied to the four spectral peaks of components I--IV in both polarizations to obtain their nominal flux densities. Next, the flux ratios of components I, II, and IV relative to component III were calculated for each polarization. The flux density of component III was then determined as the average of its left- and right-hand values. Finally, the flux densities of components I, II, and IV were obtained by multiplying the flux density of component III by their respective flux ratios in each polarization. In this way, the flux densities of all spectral components (I, II, and IV) were derived for both polarizations. Circular polarization $\Pi$ was calculated as $\Pi=(S_R-S_L )/(S_R+S_L )$, where $S_R$ and $S_L$ are flux densities of RHCP and LHCP, respectively. The measured flux density has a systematic error of up to 15\%, but the circular polarization does not contain the systematic error under the assumption of $\Pi$=0 for component III.

\section{Results}
\subsection{Spectra and circular polarization}

As an example of observational results, figures 1a and b show the 6.7 GHz methanol maser spectra of G33.641$-$0.228 observed at MJD 56893.443 (just before the burst on August 24, 2014) and MJD 56898.429 (at the burst on August 29, 2014), respectively. The solid lines represents RHCP, and the dashed lines represents LHCP. Four spectral peaks in the velocity range of 58--62 km s$^{-1}$ are displayed. Following Fujisawa et al.\ (2012, 2014b), we name these peaks as components I to IV in order of increasing velocity. Component II at Vlsr = 59.6 km s$^{-1}$ had flux densities of 22.12$\ \pm\ $0.41 Jy for RHCP and 16.55$\ \pm\ $0.44 Jy for LHCP on MJD 56893.443, while on MJD 56898.429, when the burst was detected, they became 302.21$\ \pm\ $0.39 Jy and 290.59$\ \pm\ $0.40 Jy, respectively (the errors are 1$\sigma$ thermal noise). The resulting circular polarizations are 0.144$\ \pm\ $0.022 (MJD 56893.443) and 0.0196$\ \pm\ $0.0013 (MJD 56898.429), and circular polarizations exceeding six times the thermal noise were detected. Table \ref{resulttab} shows the flux densities and circular polarizations of components I, II, and IV for these two days. No circular polarization exceeding 1.3 times the thermal noise was detected in components I and IV.

\begin{table*}[h!]
  \begin{center}
\caption{Flux densities and circular polarizations before and at the burst in 2014}
  \begin{tabular}{ccccccc}
  \hline
 MJD&&I&II&III&IV&1$\sigma$\\\hline
56893.443&RHCP (Jy)&50.8&22.1&87.0&28.9&0.41\\
Before the burst&LHCP (Jy)&51.0&16.6&&28.6&0.44\\
&Polarization&$-$0.003$\ \pm\ $0.008&0.144$\ \pm\ $0.022&-&0.005$\ \pm\ $0.015&\\\hline
56898.429&RHCP (Jy)&39.4&302.2&78.2&25.5&0.39\\
At the burst&LHCP (Jy)&38.9&290.6&&26.4&0.40\\
&Polarization&0.007$\ \pm\ $0.010&0.020$\ \pm\ $0.001&-&$-$0.019$\ \pm\ $0.015&\\\hline
  \end{tabular}
  \label{resulttab}
    \end{center}
  \label{tb1}  
\end{table*}

\begin{figure}[!htb]
 \begin{center}
  \includegraphics[width=8cm,bb= 0 0 1000 1200]{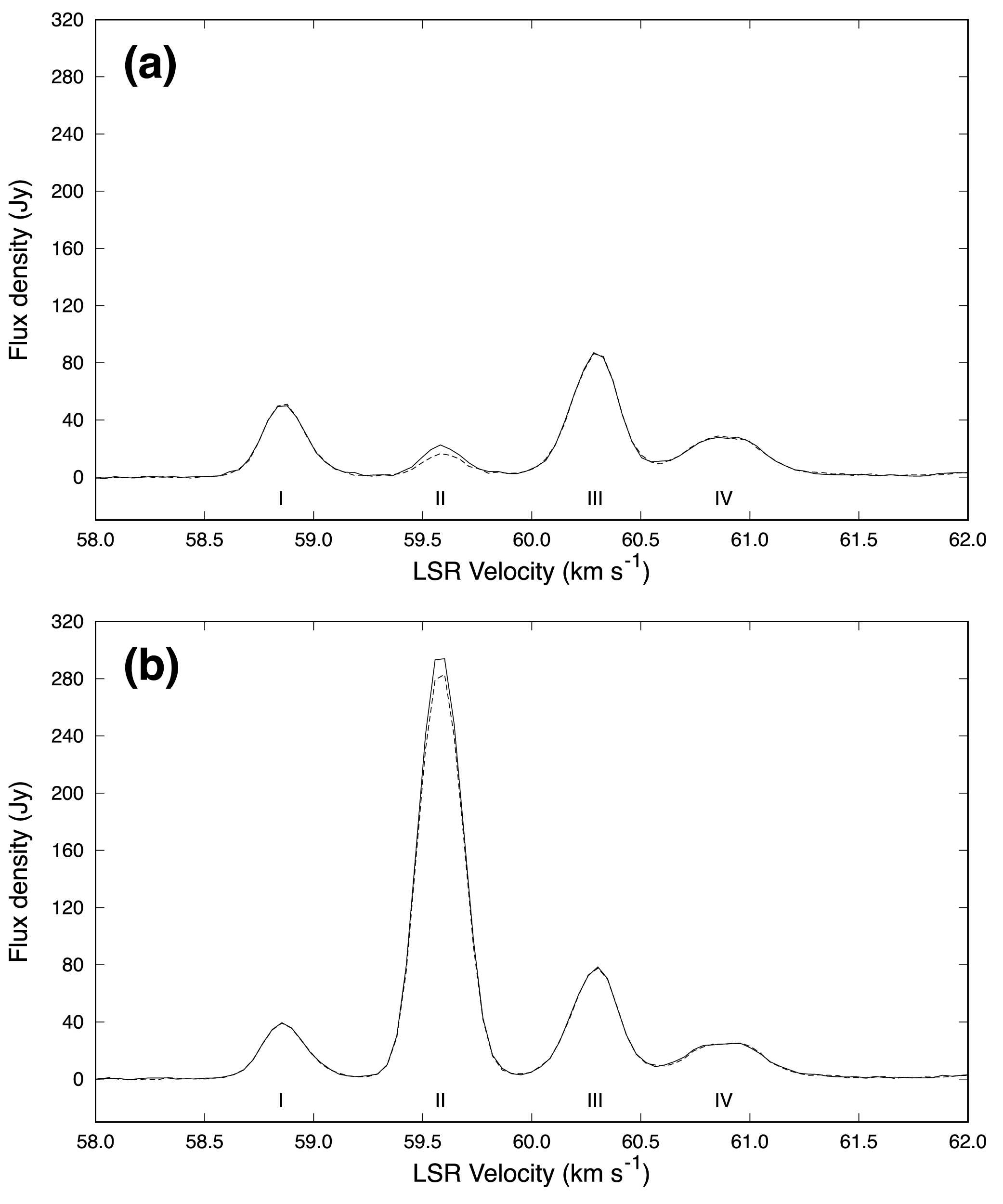}
   \caption{
   Spectra of the 6.7 GHz methanol maser in G33.641$-$0.228. RHCP is shown as a solid line and LHCP as a dotted line. (a) MJD 56893.443 (2014-8-24), before the burst, (b) MJD 56898.429 (2014-8-29), at the burst.\\
{Alt text: Two line graphs showing the 6.7 GHz methanol maser spectra of G33.641$-$0.228 befor and at the burst.}
   }
    \label{fig1}
 \end{center}
\end{figure}

\subsection{The light curve and circular polarization of component II}

\begin{figure}[!htb]
 \begin{center}
  \includegraphics[width=8.5cm,bb= 0 0 360 252]{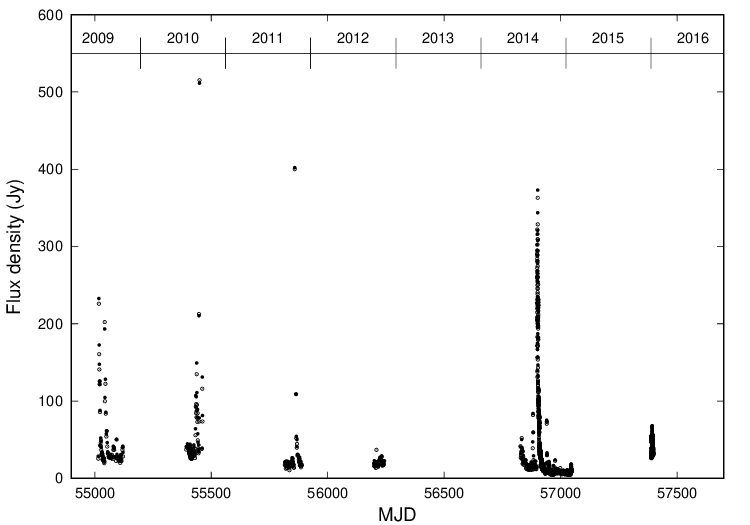}
   \caption{
   The light curves of component II over the entire observation period. Filled and open circles represent RHCP and LHCP, respectively.\\
{Alt text: Graph of flux variation of component II of the 6.7 GHz methanol maser of G33.641$-$0.228. The horizontal axis represents the observation period (2009 to 2016), and the vertical axis represents the flux density.}
   }
    \label{fig2}
 \end{center}
\end{figure}

Figure \ref{fig2} shows the light curves of RHCP (filled circle) and LHCP (open circle) of component II throughout the entire observation period. 
The light curves from 2009 to 2012 are reported in Fujisawa et al.\ (2014b). 
In 2014--2015 observation, the large burst exceeding 300 Jy occurred in August 2014 is cleary visible.

\begin{figure}[!htb]
 \begin{center}
  \includegraphics[width=8cm,bb= 0 0 360 432]{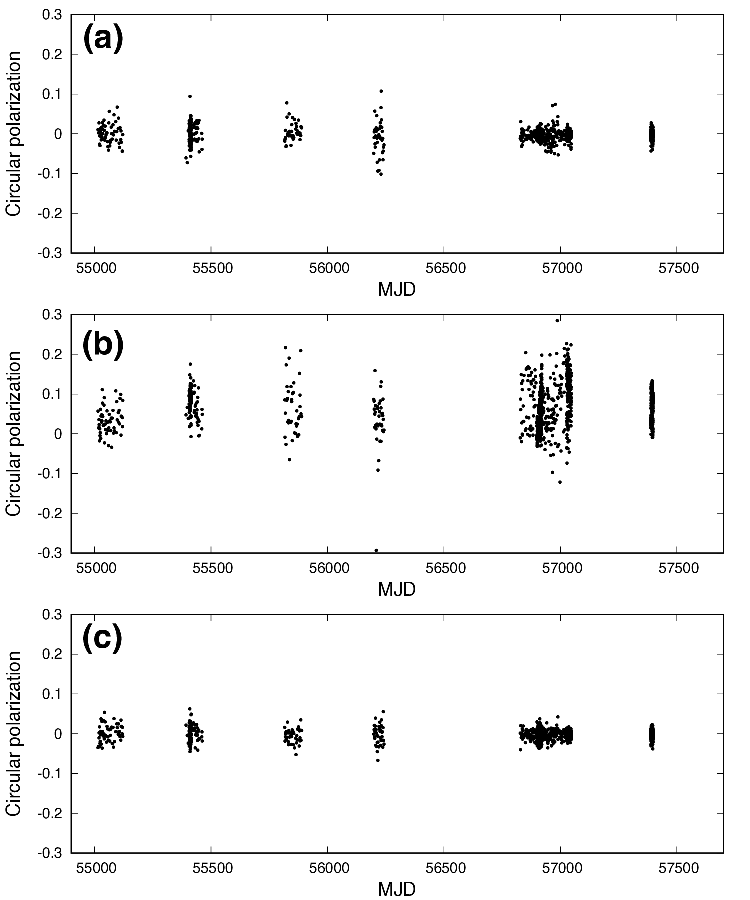}
   \caption{
  Temporal variation of circular polarization for components I, II, and IV, for panel (a), (b), and (c), respectively. Only component II shows significant positive circular polarization.\\
{Alt text: Three graphs. The horizontal axis represents the observation period (2009 to 2016), and the vertical axis represents the magnitude of the circular polarization.}
   }
    \label{fig3}
 \end{center}
\end{figure}

The circular polarizations of components I, II, and IV over the entire observation period are shown in figures \ref{fig3}a, \ref{fig3}b, and \ref{fig3}c, respectively. The mean and standard deviation of the circular polarization for all data are $-$0.003$\ \pm\ $0.017 (component I) and $-$0.002$\ \pm\ $0.014 (component IV). The circular polarizations of components I and IV are distributed around 0 throughout the entire observation period. Component II has circular polarization distributed in the range of 0.0 to 0.2.The mean and standard deviation of the polarization are 0.033$\ \pm\ $0.033 (2009), 0.079$\ \pm\ $0.034 (2010), 0.066$\ \pm\ $0.064 (2011), 0.032$\ \pm\ $0.072 (2012), 0.063$\ \pm\ $0.062 (2014--2015), 0.066$\ \pm\ $0.037 (2016), and 0.063$\ \pm\ $0.054 (all data), respectively. Significant circular polarization has been continuously observed from 2009 to 2016. A large negative circular polarization of $-$0.293 was observed only once in 2012 (MJD 56209). 

In the following subsections, we describe the light curves and circular polarization of component II in the order of the entire observation period in 2014--2015, the intensive observation during the burst period in 2014, the intensive observation in 2015, and the intensive observation in 2016.

\subsubsection{2014-2015}

\begin{figure}[!htb]
 \begin{center}
  \includegraphics[width=8cm,bb= 0 0 360 432]{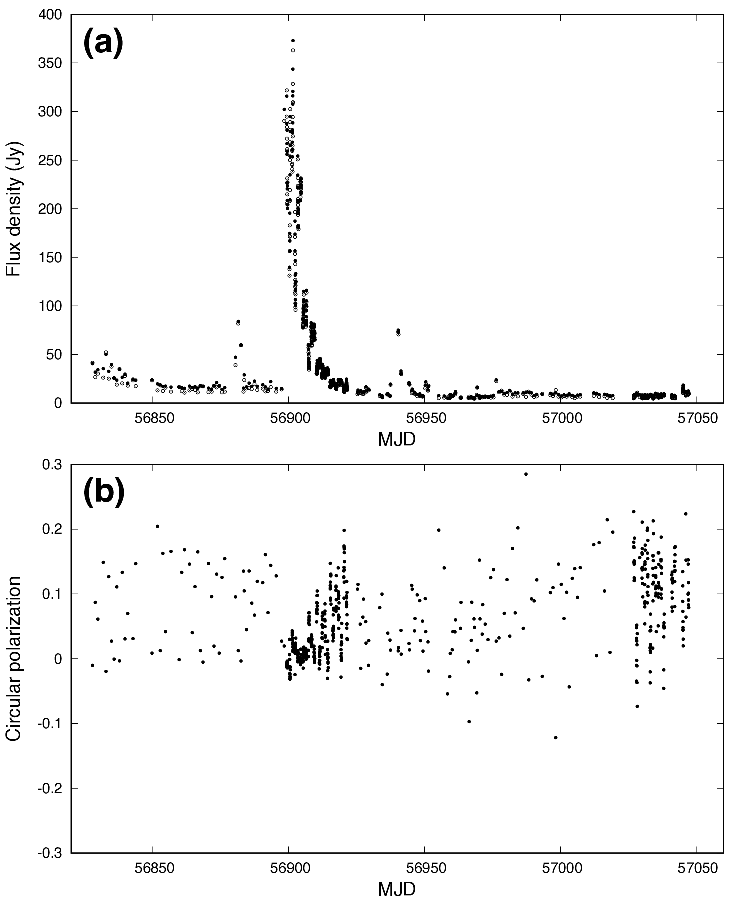}
   \caption{
  The light curves of RHCP (filled circle) and LHCP (open circle) (a) and circular polarization (b) of component II observed during 2014-2015.\\
{Alt text: Two graphs. The upper panel shows the flux variation of component II of the 6.7 GHz methanol maser of G33.641$-$0.228. The lower panel shows the circular polarization of component II. The horizontal axis is common to both the upper and lower panels and represents the observation dates from 2014 to 2015.}
   }
    \label{fig4}
 \end{center}
\end{figure}

Figures \ref{fig4}a and \ref{fig4}b show the light curve and circular polarization of component II observed during 2014--2015. The large burst occurred on MJD 56898, with smaller bursts on MJDs 56832, 56881, and 56940. The difference in flux density between RHCP and LHCP was maximum at 22 Jy (RHCP $=$ 267 Jy, LHCP $=$ 246 Jy) during the large burst on MJD 56901.375. The circular polarization shown in figure \ref{fig4}b is distributed roughly in the range of 0.0--0.2 and varies from day to day. The maximum circular polarization was 0.285$\ \pm\ $0.058 (RHCP $=$ 11.26$\ \pm\ $0.51 Jy, LHCP $=$ 6.27$\ \pm\ $0.51 Jy) observed on MJD 56987.354, which is not during a burst. During the large burst that started on MJD 56898, as the flux density increased, the circular polarization decreased towards zero. As the flux density decreased, the circular polarization increased again, approaching the distribution of non-burst periods.

\subsubsection{2014 burst}

\begin{figure}[!htb]
 \begin{center}
  \includegraphics[width=8.5cm,bb= 0 0 360 432]{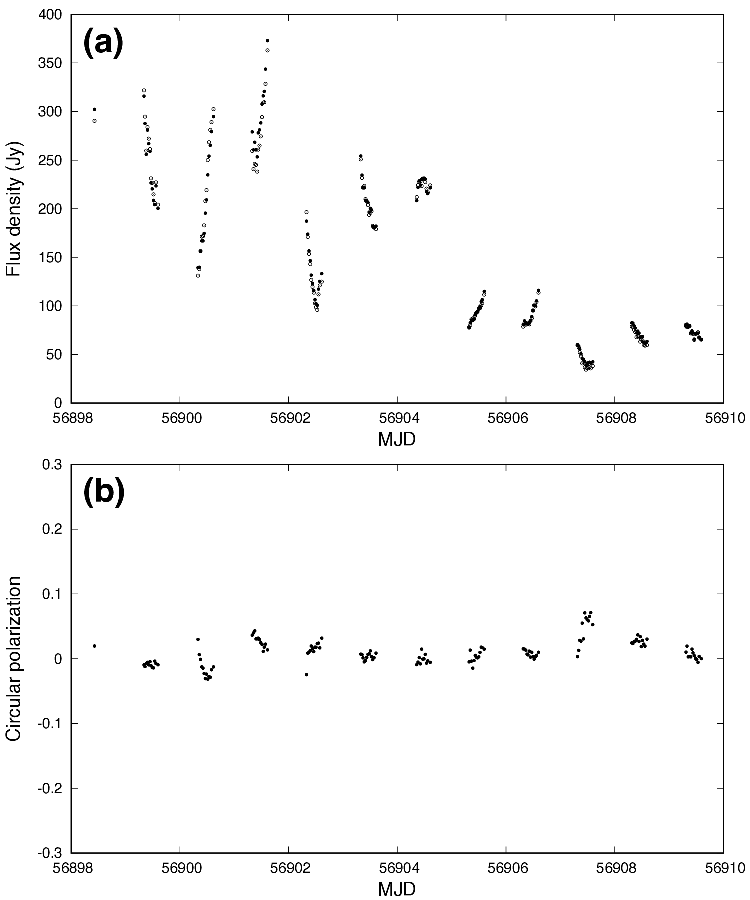}
   \caption{
The light curves of RHCP (filled circle) and LHCP (open circle) (a) and circular polarization (b) of component II observed during the large burst in 2014.\\
{Alt text: Two graphs. The upper panel shows the flux variation of component II of the 6.7 GHz methanol maser of G33.641$-$0.228. The lower panel shows the circular polarization of component II. The horizontal axis is common to both the upper and lower panels and represents the observation dates of 12 days at the burst in 2014.}}
    \label{fig5}
 \end{center}
\end{figure}

Figures \ref{fig5}a and \ref{fig5}b present the detailed light curve and circular polarization of the large burst over a 12-day period (MJD 56898--56909). The flux density repeated large increases and decreases in a short time, then gradually decreased, and became to below 50 Jy after MJD 56910. The e-folding time of the variation was as short as 0.29 d during the decrease on MJD 56902 and 0.32 d during the increase on MJD 56900. This time scale is the smallest ever observed in the variability of the 6.7 GHz methanol masers (the only report of faster flux variability is for OH masers by Clegg \& Cordes 1991). The circular polarization during this period ranged from $-$0.031 to $+$0.071 and was evidently smaller than during non-burst periods.

\subsubsection{2015 intensive monitoring}

\begin{figure}[!htb]
 \begin{center}
  \includegraphics[width=8cm,bb= 0 0 360 432]{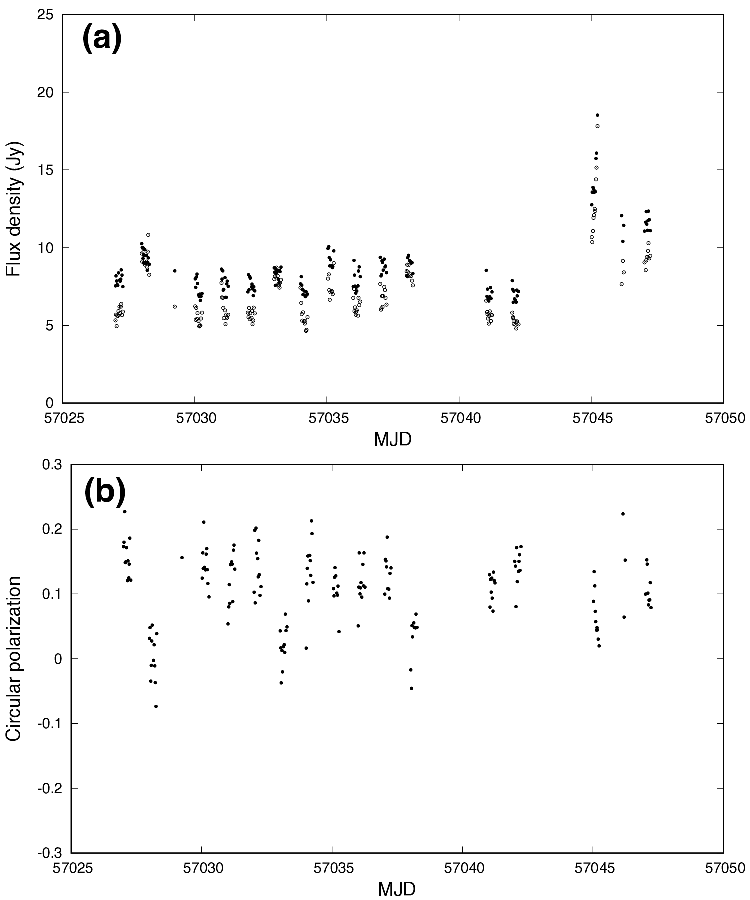}
   \caption{
The light curves of RHCP (filled circle) and LHCP (open circle) (a) and circular polarization (b) of component II observed during the intensive monitoring in 2015.\\
{Alt text: Two graphs. The upper panel shows the flux variation of component II of the 6.7 GHz methanol maser of G33.641$-$0.228. The lower panel shows the circular polarization of component II. The horizontal axis is common to both the upper and lower panels and represents the observation dates of 21 days in 2015.}
 }
\label{fig6}
 \end{center}
\end{figure}

Figures \ref{fig6}a and \ref{fig6}b show the light curve and circular polarization of the intensive observation period in 2015 (MJD 57027--57047). Except for a small burst on MJD 57045, the flux density remained in the range of 5--10 Jy. Circular polarization was distributed roughly in the range of 0.1--0.2. The errors in circular polarization were large due to the low flux density, but it is evident that the circular polarization was large and varied from day to day. For example, on MJD 57027, the average circular polarization was $0.158 \pm 0.031$, while on the following day, MJD 57028, it was $0.004 \pm 0.039$).

\subsubsection{2016 intensive monitoring}

\begin{figure}[!htb]
 \begin{center}
  \includegraphics[width=8cm,bb= 0 0 360 432]{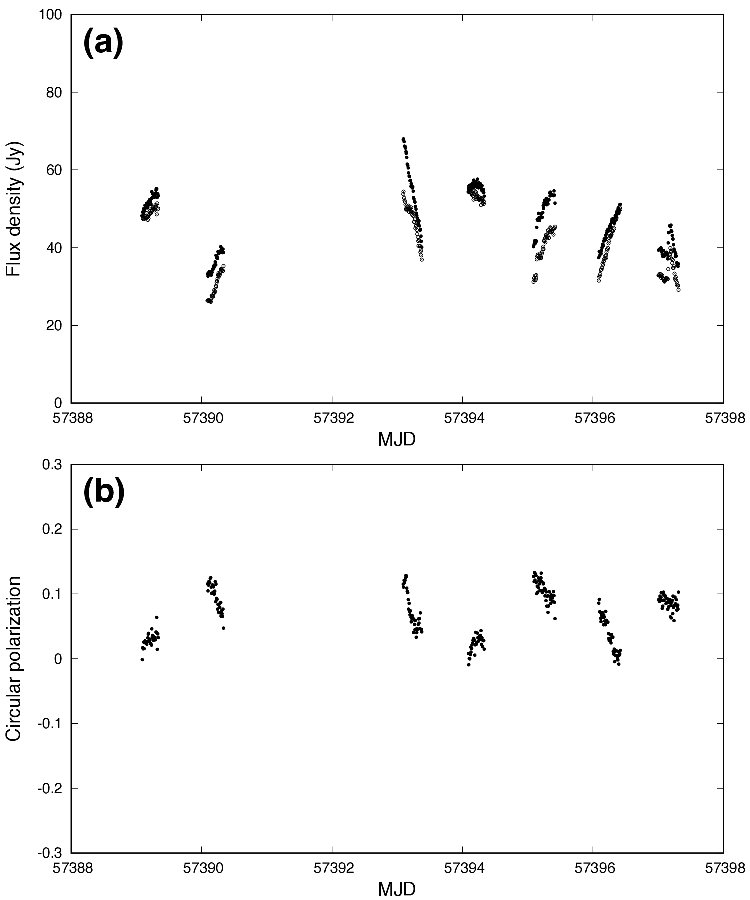}
   \caption{
The light curves of RHCP (filled circle) and LHCP (open circle) (a) and circular polarization (b) of component II observed during the intensive monitoring in 2016.\\
{Alt text: Two graphs. The upper panel shows the flux variation of component II of the 6.7 GHz methanol maser of G33.641$-$0.228. The lower panel shows the circular polarization of component II. The horizontal axis is common to both the upper and lower panels and represents the observation dates of 9 days in 2016.}
 }
\label{fig7}
 \end{center}
\end{figure}

Figures \ref{fig7}a and \ref{fig7}b present the results in 2016. At that time, the flux density of component II was relatively large, and showed a fast fluctuation. The observed range of flux density was 26--67 Jy, and the circular polarization was from $-$0.009 to $+$0.133.

\subsection{Correlation of the flux density and circular polarization}

Figure 8a shows the difference in flux density between the two polarizations (RHCP -- LHCP) plotted against the polarization-averaged flux density of component II, while figure 8b shows the corresponding circular polarization.
As the flux density increases due to bursts, the difference in flux density of the two polarizations tends to increase, while the circular polarization decreases and approaches to 0.

Figure \ref{fig9} shows the distribution of the circular polarization to the flux density for 7 d in the 2016 intensive monitoring to demonstrate the correlation of the flux density and the circular polarization. Table \ref{tab3} shows the correlation between the flux density and the circular polarization for each observation day. The correlation varies from day to day. On January 3, 8, and 9, circular polarization decreased as flux density increased, whereas on January 6 it showed a positive correlation. The correlation was weak on January 2, 7, and 10. Overall, no clear trend was found in the relationship between flux density and circular polarization during this period.

\begin{figure}[!htb]
 \begin{center}
  \includegraphics[width=8cm,bb= 0 0 1000 1201]{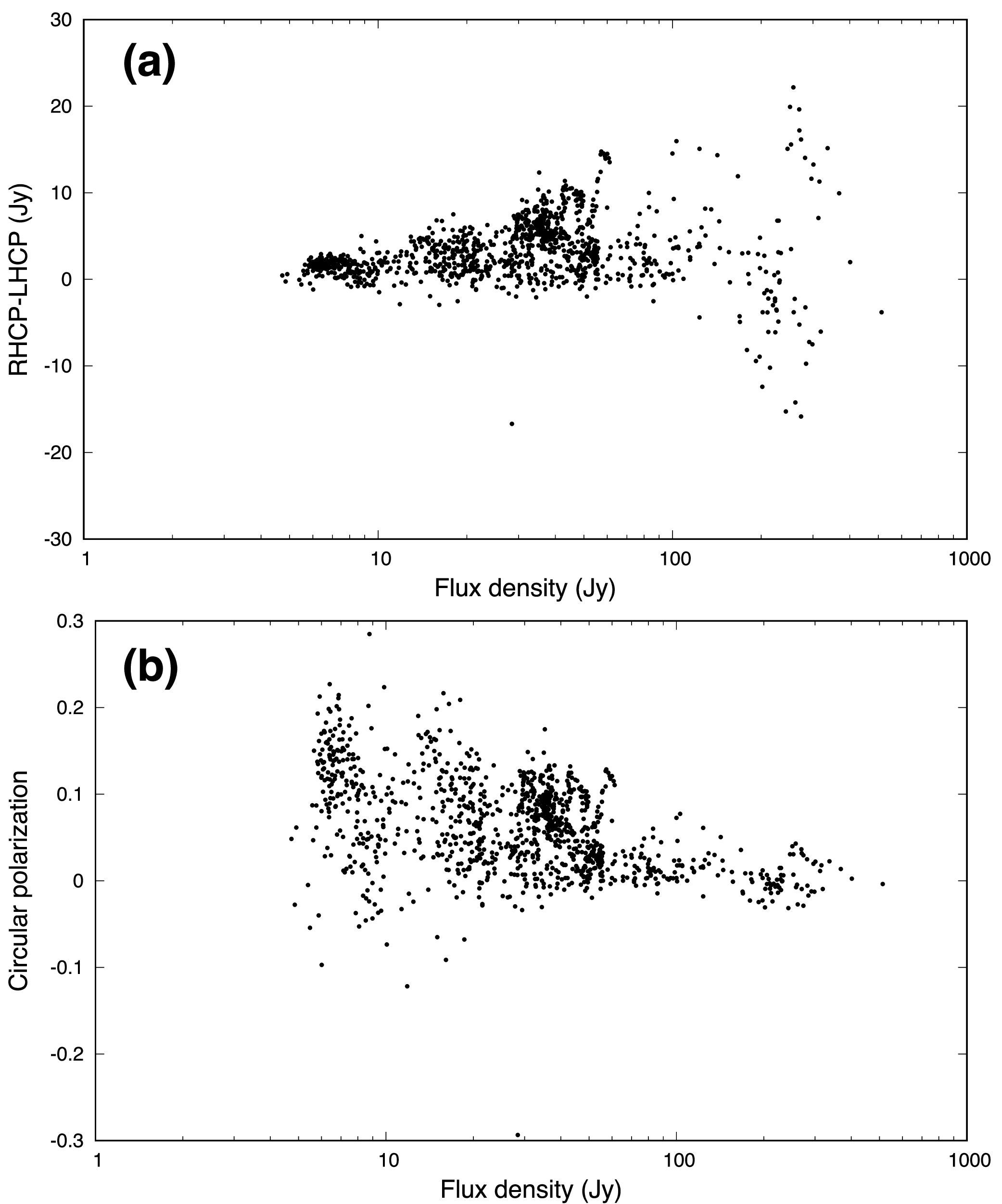}
   \caption{
(a) the difference of the flux density of two polarization (RHCP-LHCP) to the average flux density, (b) the circular polarization to the average flux density of component II.\\
{Alt text: Two graphs. The vertical axis of the upper panel shows the difference in flux density between RHCP and LHCP for component II of the 6.7 GHz methanol maser of G33.641$-$0.228. The vertical axis of the lower panel shows the circular polarization of component II. The horizontal axis for both panels shows the flux density of component II (average of RHCP and LHCP).}
 }
\label{fig8}
 \end{center}
\end{figure}

\begin{figure}[!htb]
 \begin{center}
  \includegraphics[width=8cm,bb= 0 0 360 252]{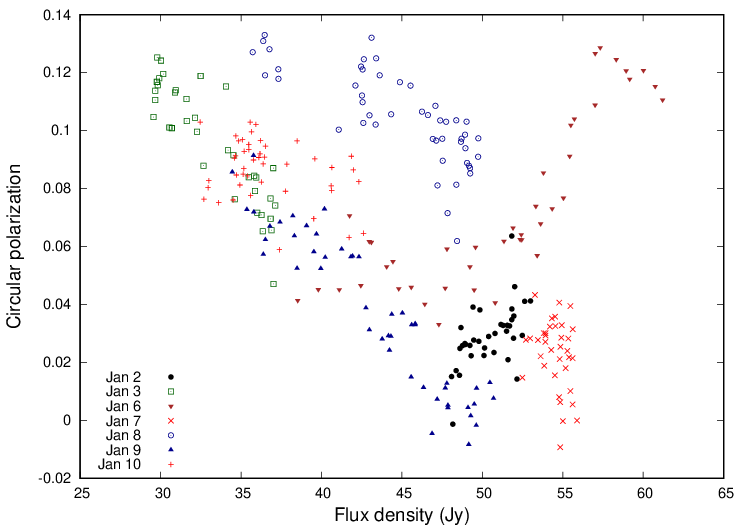}
   \caption{
Circular polarization plotted against to the polarization averaged flux density in 2016 intensive observation.\\
{Alt text: The vertical axis shows the circular polarization of component II of the 6.7 GHz methanol maser of G33.641$-$0.228. The horizontal axis shows the flux density of component II (average of RHCP and LHCP).}
 }
\label{fig9}
 \end{center}
\end{figure}

\begin{table*}[h!]
  \begin{center}
\caption{Correlation of the flux density and circular polarization in 2016 intensive monitoring}
  \begin{tabular}{ccccccc}
  \hline
 Date of 2016&MJD&Duration&Number of scan&Average flux density&Average polarization&Correlation coefficient\\
       &      &(hr)&  &(Jy) &      &       \\\hline
Jan. 2 &57389 &6  &36 &50.5 &0.029 &0.56   \\
Jan. 3 &57390 &6  &36 &33.2 &0.096 &$-$0.88\\
Jan. 6 &57393 &7  &42 &50.9 &0.073 &0.82   \\
Jan. 7 &57394 &6  &36 &54.5 &0.023 &$-$0.29\\
Jan. 8 &57395 &8  &48 &44.8 &0.105 &$-$0.77\\
Jan. 9 &57396 &8  &48 &43.2 &0.038 &$-$0.94\\
Jan. 10&57397 &7.5&45 &36.7 &0.087 &$-$0.33\\\hline
  \end{tabular}
  \label{tab3}
    \end{center}
  \begin{tabnote}
  \end{tabnote}
\end{table*}

\subsection{Negative circular polarization}

\begin{figure}[!htb]
 \begin{center}
  \includegraphics[width=8cm,bb= 0 0 360 432]{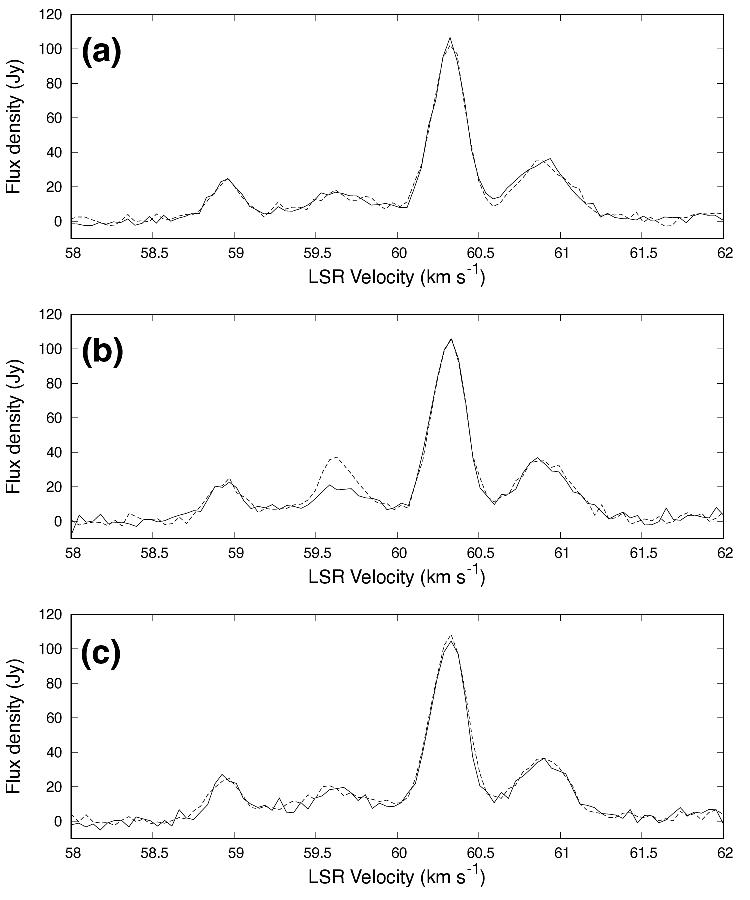}
   \caption{
Spectra on October 8, 2012 (MJD 56208.531, top), 9 (MJD 56209.528, middle), and 11 (MJD 56211.523, bottom).\\
{Alt text: Three graphs showing the 6.7 GHz methanol maser spectrum of G33.641$-$0.228 obtained in three consecutive observations.}
 }
\label{fig10}
 \end{center}
\end{figure}

On October 9, 2012, a large negative circular polarization of $-$0.293$\ \pm\ $0.076 was observed. Figures \ref{fig10}a, \ref{fig10}b, \ref{fig10}c show the spectra from the day before (MJD 56208.531), the day (MJD 56209.528), and two days after (MJD 56211.523) the event. In the spectrum on MJD 56209, the LHCP of component II is obviously larger than RHCP. This variability occured in timescale of less than a day and returned to its original level in less than two days. The flux density and circular polarization of component II over the three days are shown in Table \ref{tab4}.

\begin{table}[h!]
  \begin{center}
\caption{ Flux densities and circular polarizations in days of October 8, 9 and 11, 2012 (MJD 56208--56211)}
  \begin{tabular}{cccc}
  \hline
 MJD&RHCP&LHCP&Polarization\\
    &(Jy)&(Jy)&\\\hline
56208.531&16.6$\ \pm\ $2.2&16.2$\ \pm\ $2.0&0.012$\ \pm\ $0.127\\
56209.528&20.1$\ \pm\ $2.2&36.8$\ \pm\ $2.2&$-$0.293$\ \pm\ $0.076\\
56211.523&18.2$\ \pm\ $2.1&18.7$\ \pm\ $1.1&$-$0.014$\ \pm\ $0.112\\\hline  
\end{tabular}
  \label{tab4}
    \end{center}
  \begin{tabnote}
  \end{tabnote}
\end{table}

\begin{figure*}[!htb]
 \begin{center}
  \includegraphics[width=16cm,bb= 0 0 1152 406]{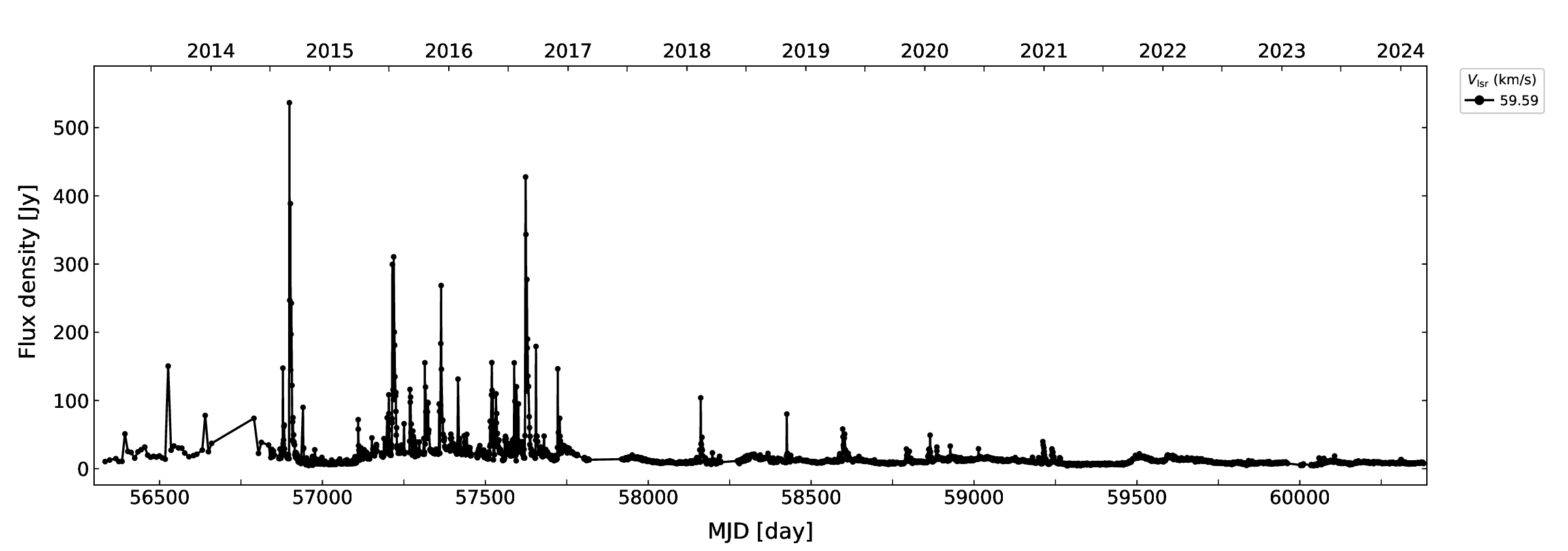}
   \caption{
The light curve of component II from MJD 56299 to 60380.\\
{Alt text: The vertical axis represents the flux density of component II of the 6.7 GHz methanol maser of G33.641$-$0.228, and the horizontal axis represents the observation period from 2014 to 2024.}
 }
\label{fig11}
 \end{center}
\end{figure*}

\subsection{Properties of the flux and circular polarization variability}
The flux variability and circular polarization properties obtained from the observation can be summarized as follows:

\begin{itemize}
      \item Flux variability: During the large burst in 2014, a maximum flux density of up to 350 Jy was observed, and a rapid rise and fall of flux density occurred repeatedly, eventually decreasing to 50 Jy or less after MJD 56910, 12 d later the burst start. The time scale of the variability (e-folding time) was as short as 0.29 d during the decline on MJD 56902 and 0.32 d during the rise on MJD 56900.
      \item Components I and IV are not polarized. Their polarizations are 0 within the error.
      \item Component II shows positive circular polarization ranging from 0.0 to 0.2 in most cases. The polarization varies on a time scale of less than one day.
      \item The circular polarization does not match the flux density variability. When the large burst occured in 2014 and the flux density increased, the polarized flux density also increased, but the polarization decreased to 0. On the other hand, 2016 intensive monitoring showed no systematic correlation of circular polarization and flux density.
      \item These properties have been observed continuously from 2009 to 2016.
      \item Only once, a large negative circular polarization of $-$0.293 was observed.
\end{itemize}

\section{Discussions}

In the data reduction, we have assumed that the circular polarization of component III is zero. The derived circular polarization of components I and IV based on this assumption was zero within the error. This does not prove but supports the assumption that the circular polarization of component III is zero.

Circularly polarized radiation has been detected only in component II, and the bursts also occur only in this component. Therefore, the burst and the circularly polarized radiation could be related and/or caused by a common physical mechanism. Linear polarization often arises from interstellar matter along the propagation path. If dust and magnetic fields exist along the propagation path of visible light from stars, linear polarization is generated by the alignment of the dust (e.g., Heiles\ 2000). This mechanism, however, is unlikely to produces circular polarization during the propagation of 6.7 GHz radio emission.

If a strong magnetic field exists in the maser emission region, circular polarization can be generated. The Zeeman effect produces a frequency difference between the RHCP and LHCP. Vlemmings\ (2008) observed the line-of-sight velocity difference of $\Delta v = -$0.89$\ \pm\ $ m s$^{-1}$ and he derived the magnetic field strength of 18 mGauss for G33.641$-$0.228 with a velocity shift of $\Delta v = -$49 m s$^{-1}$ Gauss$^{-1}$. Takagi et al. (2021) derived the Zeeman splitting coefficient as $0.096 \pm 0.005$ Hz mGauss$^{-1}$ ($\Delta v = -$4.3 m s$^{-1}$ Gauss$^{-1}$), about ten times smaller than the value adopted by Vlemmings (2008). Using this coefficient, the magnetic field of G33.641$-$0.228 is estimated to be 205 mG. The magnetic field strength is comparable to that of methanol masers in other sources (Vlemmings 2008), and it is unlikely that strong circular polarization due to the magnetic field occurs only in this component of only in this source.

It is well established that magnetic fields are involved in burst-like phenomena in YSOs. Takasao et al. (2019) conducted three-dimensional magnetohydrodynamic simulations to explain X-ray flares in YSOs and showed that the flares are caused by magnetic reconnection. Although X-ray flares differ from the maser bursts in G33.641$-$0.228, magnetic fields may likewise play a role in the maser bursts.

When circularly polarized emission occurs behind a maser cloud, the amplified maser emission also become circularly polarized. The circular polarization observed in G33.641$-$0.228 may have been amplified by the circularly polarized continuum radiation behind it. There are two well-known mechanisms for circularly polarized continuum radiation. One is gyro-synchrotron radiation observed in solar radio bursts (Dulk\ 1985), where high-energy charged particles interact with the magnetic field and emit circularly polarized radiation. The other is gyro-resonance radiation, in which thermal plasma within a strong magnetic field emits circularly polarized radiation (Shibasaki et al.\ 1994). In YSOs, circularly polarized radiation has been observed during flares of T Tauri stars in the microwave band (Skinner \& Brown\ 1994; Phillips et al.\ 1996).

We adopt a working hypothesis that an explosive event similar to solar radio burst occurred on the surface of the protostar and the circularly polarized continuum radiation was generated, then the radiation was amplified by the maser cloud of component II on the line of sight from the explosive event region to the observer. In this hypothesis, the flux variability of the maser burst is not due to the variability of the physical property of the maser cloud, but due to the variability of the background radiation. The explosive event region of the protostar, the component II maser cloud, and the observer are aligned linearly, while other maser clouds are not. Based on this working hypothesis, we attempt to explain the observed properties of the burst and the circular polarization in the following.
\begin{itemize}
\item Property 1: Short-term flux variation. The rise and fall of the flux, repeating with short intervals, are observed in the solar radio bursts (microwave IV bursts). The timescale of solar radio bursts is 0.01 d (Dulk\ 1985). The time scale of the flux variability of the maser bursts is 30 times larger than that of solar radio bursts, but other qualitative properties are very similar. Note that radio bursts with circular polarization have also been observed in the low-mass pre-main-sequence stars T Tauri and V773 Tau, with variability timescales of less than a few days for T Tau and a few hours for V773 Tau (Skinner \& Brown\ 1994; Phillips et al.\ 1996). 

\item Property 2: Circular polarization and its variability. In solar radio burst, circular polarization due to either gyro-synchrotron radiation or gyro-resonance radiation is observed and fluctuates over short periods of time. The circular polarization degree is about 10\% for gyro-synchrotron radiation (Dulk\ 1985). If a similar event occurred in G33.641$-$0.228, the observed fluctuations in circular polarization degree can be explained.

\item Property 3: The correlation between flux density and polarization changes from day to day (figure \ref{fig9}) for the maser burst. During the large burst, circularly polarized flux increased, but polarization decreased (figure \ref{fig8}). There is no simple explanation for these properties.

\item Property 4: The circular polarization is present during the quiescent period, and the polarization degree is high. It is possible that there is a circular polarization radiation component that was not directly related to the burst, but it is unclear why only component II of this source shows burst and circularly polarized radiation.

\item Property 5: Positive circular polarization of 0.0--0.2 from 2009 to 2016 has been continuously observed. 
According to the working hypothesis, there is a magnetic field with a global structure in the explosive event region, and the direction of the magnetic field in the region is sustained for 7 years.

\item Property 6: On October 9, 2012, a strong negative polarization of $-0.293$ was observed. This requires an ad hoc explanation that, on that day, a reversed circular polarization coincidentally occurred on the surface of the protostar.

\end{itemize}

Taken together, the working hypothesis can explain the generation of circular polarization and short-term flux variability, while it does not well explain the correlation between flux density and polarization, the presence of strong circular polarization radiation even in the quiet period, the stability of positive circular polarization for 7 years, and the occurrence of a large negative circular polarization observed only once.

While our model is a qualitative hypothesis to explain the phenomenon, Khaibrakhmanov et al. (2025) propose a physical model called reconnection in the magnetosphere of a protostar. Their model is as follows: Magnetic reconnection occurs at the point where the magnetosphere of the protostar meets the magnetic field of the disk, releasing the magnetic field energy. The released energy locally heats the gas that produces masers. This changes the excitation state, causing a maser burst. This model can well explain the nature of the observed short-term variability, but the circular polarization is not discussed by Khaibrakhmanov et al. (2025).

If VLBI observations will be made during bursts, and the location where the bursts occur will be identified, leading to an understanding of the burst mechanism. Additionally, if the background continuum emission is amplified by masers, the detection of this continuum emission would be a clue. It is expected that the continuum emission is circularly polarized and shows the same light curve with the maser, and that the continuum emission region exists only behind the component II maser cloud and not behind the other maser clouds. The size of the continuum emission region could be estimated from the timescale of the variability of 0.3 d, which would be smaller than 50 au or 13 mas and can be easily measured by VLBI. If such continuum emission will be detected, the burst mechanism can be clarified, consequently the amplification rate of the maser can be directly measured.

Fujisawa et al. (2012) and Bartkiewicz et al. (2009, 2012) reported VLBI observations of the 6.7 GHz methanol maser in G33.641$-$0.228, but these observations were not carried out during the bursting period. Unfortunately, no burst was observed after 2022 by the long term methanol maser monitoring program of Hitachi 32-m telescope\footnote{Ibaraki 6.7 GHz class II methanol maser database (iMet)} as shown in figure \ref{fig11}.

\section{Conclusion}

The 6.7 GHz methanol maser in the high-mass star-forming region G33.641$-$0.228 shows burst-like flux variability. We have made monitoring observations to investigate the mechanism of this burst with the Yamaguchi 32-m radio telescope from 2009 to 2016. The data were analyzed independently for both left and right circular polarization to examine the circular polarization property of the burst. 

Through this observation, the following facts have been revealed. After the large burst, a short-term rise and fall of flux density with a time scale (e-folding time) of 0.3 d were observed. Among the spectral components, only component II, which showed the burst, exhibited large circular polarization. Circular polarization was observed from 2009 to 2016, and it ranged roughly from 0.0 to 0.2 and varied on a time scale of less than one day. When a burst occurred, the circularly polarized flux increased as the averaged flux density, but the circular polarization tended to decrease when the flux density increased. The variability of the flux density and the circular polarization did not show any clear correlation at the low flux density period. Only once a large negative circular polarization of $-$0.293 was observed during the observation period.

We proposed a model to explain these properties, in which an explosive event similar to a solar radio burst occurs on the surface of the protostar and the generated radiation with high circular polarization is amplified by the component II maser cloud on the line of sight from the location of the explosive event to the observer. This model can explain the generation of large circular polarization and short-term flux variations. Khaibrakhmanov et al. (2025) proposed a physical model of reconnection between the protostellar magnetosphere and the disk magnetic field to explain the observed short-term flux variability.

\begin{ack}
This work is partially supported by the Inter-university collaborative project, the Japanese VLBI Network (JVN) of the National Astronomical Observatory of Japan. The authors thank to Dr.\ Hachisuka, Dr.\ Matsumoto, Mr.\ Takase, Ms.\ Fukui, and Ms.\ Koga for their assistance with the observations. KF thanks to Dr. Kazi Rygl and Dr. Yuanwei Wu for providing the distance information. The authors also thank to KDDI Corporation for their support of the Yamaguchi 32-m radio telescope.

\end{ack}

{}

\end{document}